# Symmetric drainage flow of a compressible fluid from a fracture: analytical solution and slip-like flow rate

Di Shen  and  Kang Ping Chen*

School for Engineering of Matter, Transport and Energy

Arizona State University

Tempe, AZ 85287-6106

## Abstract

Symmetric drainage flow of a compressible fluid from a fracture modeled as a long narrow channel is studied on the basis of linearized compressible Navier-Stokes equations with no-slip condition. The Helmholtz decomposition theorem is used to decompose the velocity field into an irrotational part and a solenoidal part. The irrotational velocity is driven by the fluid's volumetric expansion whilst the role of the solenoidal velocity is to enforce the no-slip condition for the overall velocity and it does not contribute to the mass flow rate. It is found that at large times this no-slip flow exhibits a time-dependent slip-like mass flow rate linearly proportional to the channel gap instead of the cubic power of the gap from the Poiseuille-type of flow. The drainage rate is also proportional to the kinematic viscosity, opposite to Poiseuille-type of flow which has a drainage rate proportional to the inverse of the kinematic viscosity. The same drainage rate formula also applies to drainage flow from a semi-sealed microchannel.

*Correspondence to:  k.p.chen@asu.edu.

1. **Introduction**

It is well known that when a gas moves steadily through a microchannel it can exhibits a significantly higher mass flow rate than that predicted by the Poiseuille's law (Knudsen 1909; Arkilic 1994; Harley et al. 1995; Shih et al. 1996; Arkilic et al 1997a, b; Jang et al. 2002; Maurer et al. 2003; Colin 2005; Ewart et al. 2006, 2007). This departure from the classical result has been often attributed to a breakdown of the continuum assumption which is the foundation of the Navier-Stokes equations. One cause of such a breakdown is gas rarefaction which can make the mean free path of the gas comparable to the characteristic length of the confining geometry (Tsien 1946; Karniadakis et al. 2005; Colin 2005). In many studies of microchannel flow in the so-called slip flow regime, the Navier-Stokes equations are presumably still valid and the no-slip condition on the channel wall is replaced by a slip condition (Maxwell 1879; Beskok & Karniadakis 1999; Zohar et al. 2002; Hadjiconstantinou 2003, 2006; Wu 2008; Veltzke & Thaming 2012; Zhang et al. 2012). The results from this approach matched the observed mass flow rate when a slip parameter is adjusted to a suitable value. Another approach is the Extended Navier-Stokes Equations (ENSE) formulation which includes an additional diffusive mass flux (or volume flux) from self-diffusion caused by a local density gradient, which is allowed to slip on the wall (Brenner 2005; Durst et al. 2006; Chakraborty & Durst 2007; Dongari et al. 2007; Dongari et al. 2009; Dadzie & Brenner 2012; Jaishankar & McKinley 2014). A consensus formed from these studies is that in order to match the elevated mass flow rate observed in the slip flow regime, the Navier-Stokes equations must be supplemented with a slip-type boundary condition; or the governing equations must incorporate additional slip-like terms such as those introduced in the ENSE.

Unsteady viscous compressible flow in a microchannel can exhibit quite different flow behaviors even when the no-slip condition is enforced. One well-known example is acoustic streaming (Squires 2005; Friend & Yeo 2011). The mean (period-averaged) mass flow rate in an acoustic streaming flow is constant in time but is slip-like as it is proportional to the square of the tube radius instead of the quartic power. However, acoustic streaming requires inertia and inertia is quite small for micro-conduits. Recently, we have found that a slip-like mass flow rate also occurs in the absence of inertia during the drainage flow of a viscous compressible fluid from a small capillary (Chen & Shen, 2018a, b). The mass flow rate for such a drainage flow is proportional to the kinetic viscosity of the fluid and the square of the tube radius, even though the velocity satisfies the no-slip condition on the capillary wall. This result has been used to explain the phenomenon of shale gas and shale oil production from the nanopores of shale matrix in hydraulically fractured reservoirs (Chen & Shen, 2018 b). In the present paper, we extend our earlier work by presenting an analytical solution to the classical compressible Navier-Stokes equations (CNSE) with no-slip condition for a microchannel flow. Drainage flow in this geometry is also motivated by the phenomenon of fluid production from massively hydraulically fractured shale formations (Hill & Nelson 2000; Nelson 2009; Hughes 2013). There are abundant small natural fractures in shale formation and they store natural gas and oil, just like the nanopores in the shale matrix. These natural fractures can be modeled as long and narrow microchannels. Drainage flow of the compressed fluid from these natural fractures is driven by the fluid's volumetric expansion when the ends of the fracture (channel) are opened to a lower pressure (or density) environment. As reported below, the solution to the linearized compressible Navier-Stokes equations with no-slip condition for this unsteady flow exhibits a slip-like mass flow rate, similar to the case of a circular capillary (Chen & Shen, 2018a, b). No additional hypothesis is introduced in the analysis; and the presented analytical solution is valid even as the channel gap tends to zero, provided the Navier-Stokes equations are still applicable.

We begin the paper with a mathematical description of the symmetric drainage flow problem, followed by an analytical solution to the initial-boundary-value-problem for the density perturbation. The key to the solution of the velocity field is to utilize the Helmholtz decomposition theorem for the velocity. Mechanism of fluid production will be discussed as well.

## 2. Mathematical formulation

We study symmetric drainage flow of a single phase compressible fluid from a long and narrow channel (Fig. 1). The channel has a narrow gap $2H$ and a length $2L$ with $L >> H$. The compressed fluid stored in the channel is initially at rest with a density $\rho_i$ with both ends closed. The channel is embedded in a bath of the same compressible fluid maintained at a density $\rho_e$ slightly lower than $\rho_i$. At $t=0$, the two ends of the channel at $x=-L, L$ are opened fully, and the densities at the exits are maintained at $\rho_e$ at all times thereafter. Volumetric expansion causes the fluid to drain from the microchannel through the ends symmetrically. The density difference $\rho_i - \rho_e$ is assumed to be small and temperature variation is negligible. The Mach number defined as the ratio between a characteristic fluid velocity and the speed of sound is assumed to be small; as such the continuity and the compressible Navier-Stokes equations (Chorin & Marsden 1992) can be linearized around the final equilibrium state $(\rho, \mathbf{v}) = (\rho_e, 0)$ so that (Morse & Ingard 1968; Temkin 1981)

$$\frac{\partial \rho'}{\partial t} + \rho_e \nabla \cdot \mathbf{v}' = 0, \tag{1}$$

$$\rho_e \frac{\partial \mathbf{v}'}{\partial t} = -\nabla p' + \left(\mu_b + \frac{1}{3}\mu\right)\nabla(\nabla \cdot \mathbf{v}') + \mu \nabla^2 \mathbf{v}', \tag{2}$$

where the density perturbation $\rho' = \rho - \rho_e$; $p'$ is the pressure perturbation; $\mathbf{v}'$ is the velocity perturbation which is assumed to be symmetric about the channel centerline $y = 0$; $\mu, \mu_b$ are the shear and bulk viscosities of the fluid, respectively. Eqns. (1) and (2) can be used to derive a damped wave equation for the density perturbation (Morse & Ingard 1968)

$$\frac{\partial^2 \rho'}{\partial t^2} = \left(c^2 + D_\rho \frac{\partial}{\partial t}\right)\nabla^2 \rho', \tag{3}$$

where the diffusion coefficient $D_\rho = (\mu_b + 4\mu/3)/\rho_e$ characterizes the diffusion of a small density disturbance and $c$ is the speed of sound. Eqn. (3) is decoupled from the velocity field. It must be emphasized that eqn. (3) is not an *ad hoc* equation: it is as rigorous as the linearized system of equations (1) and (2). The boundary and initial conditions for the density perturbation are:

$$x = 0: \frac{\partial \rho'}{\partial x} = 0; \; x = L: \rho' = 0; \tag{4}$$

$$t = 0: \rho' = \rho_i - \rho_e; \frac{\partial \rho'}{\partial t} = 0. \tag{5}$$

The boundary condition on $x = 0$ is due to symmetry. The density at the exit is fixed to $\rho_e$ so that $\rho' = 0$ at the exit. Justification of using this exit boundary condition for drainage flow has been shown by Chen & Shen (2018a). Unlike the velocity, the boundary value of the density on the channel wall does not need to be specified (Chen & Shen, 2018a, b). Of particular interest is that the mass flow rate at the exit $\dot{m}_e(t)$ can be computed once the instantaneous density distribution is obtained without the need for explicit knowledge of the corresponding velocity field, as the integral of the continuity equation over one half of the channel (due to symmetry) gives

$$\dot{m}_e(t) = - \int_{\text{half channel}} \frac{\partial \rho'}{\partial t} d\text{v}. \tag{6}$$

The velocity field can be computed once the density distribution for the linearized flow is obtained. From the Helmholtz decomposition theorem (Aris 1989; Leal 2010; Panton 2013), the velocity field $\mathbf{v}'$ can be decomposed into the sum of an irrotational part $\mathbf{v}'_{IR}$ and a rotational part $\mathbf{v}'_{RT}$,

$$\mathbf{v}' = \mathbf{v}'_{IR} + \mathbf{v}'_{RT}, \tag{7}$$

where the irrotational part is a potential flow and the rotational part is solenoidal (incompressible):

$$\nabla \times \mathbf{v}'_{IR} = 0, \quad \mathbf{v}'_{IR} = \nabla \Phi, \tag{8}$$

$$\nabla \cdot \mathbf{v}'_{RT} = 0. \tag{9}$$

In the above, $\Phi$ is the scalar velocity potential for the irrotational flow. These two parts of the velocity field are called the longitudinal mode and transverse mode respectively in the acoustic literature (Morse & Ingard 1968; Temkin 1981). In the linear theory, the two modes are only coupled through the no-slip condition on the channel wall for the total velocity $\mathbf{v}'$,

$$y = \pm H : \mathbf{v}' = \mathbf{v}'_{IR} + \mathbf{v}'_{RT} = 0. \tag{10}$$

The irrotational velocity can be determined from the continuity equation which now takes the form

$$\frac{\partial \rho'}{\partial t} + \rho_e \nabla \cdot \mathbf{v}'_{IR} = 0, \tag{11}$$

since $\nabla \cdot \mathbf{v}'_{RT} = 0$. At $x = 0, \mathbf{v}'_{IR} = 0$. The solenoidal velocity $\mathbf{v}'_{RT}$ is governed by the equations for an incompressible flow, which can be expressed in terms of the incompressibility condition and the linearized vorticity diffusion equation

$$\nabla \cdot \mathbf{v}'_{RT} = 0, \tag{12}$$

$$\frac{\partial \boldsymbol{\omega}}{\partial t} = \nu_e \nabla^2 \boldsymbol{\omega}, \tag{13}$$

$$\boldsymbol{\omega} = \nabla \times \mathbf{v}' = \nabla \times \mathbf{v}'_{RT}, \tag{14}$$

where the kinematic viscosity $\nu_e = \mu / \rho_e$. The solenoidal velocity is solely driven by the no-slip condition imposed on the overall velocity, equation (10).

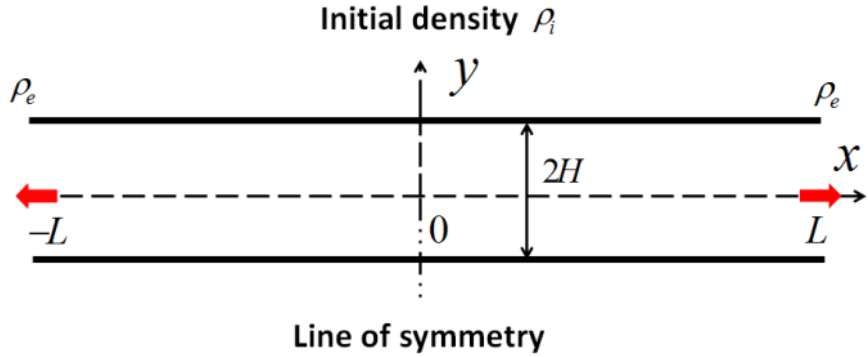

**Fig. 1** Schematic of symmetric drainage flow from a channel. The initial density is $\rho_i$; and the exit density is $\rho_e$.

## 3. Acoustic waves and transverse density relaxation in a narrow channel

The initial-boundary-value-problem for the density perturbation (3)-(5) can be solved using the method of separation of variables (Chen & Shen, 2018a, b). The complete solution for $\rho'(x, y, t)$ is given by

$$\rho'(x,y,t) = \frac{4(\rho_i - \rho_e)}{\pi} \left[ \sum_{n=0}^{N_d} \frac{(-1)^n}{2n+1} \exp[-\gamma_n t] \cos \omega_n t \cos \frac{(2n+1)\pi x}{2L} \right. \\ \left. + \sum_{n=N_d+1}^{\infty} \frac{(-1)^n}{2n+1} \exp[-\gamma_n t] \cosh \gamma_n \sqrt{1-\varepsilon_n} t \cos \frac{(2n+1)\pi x}{2L} \right] \\ + RE \sum_{n=0}^{\infty} \sum_{m=1}^{\infty} A_{nm} \cos(\alpha_m y) \left( \exp[K_{nm}^+ t] - \exp[K_{nm}^- t] \right) \cos \frac{(2n+1)\pi x}{2L}, \quad (15)$$

where "$RE$" stands for the real part and

$$\gamma_n = \frac{D_\rho}{2} \frac{(2n+1)^2 \pi^2}{4L^2}, \quad \omega_n = c \frac{(2n+1)\pi}{2L} \sqrt{1 - \frac{(2n+1)^2 \pi^2 D_\rho^2}{16 c^2 L^2}},$$

$$\varepsilon_n = \frac{16 c^2 L^2}{(2n+1)^2 \pi^2 D_\rho^2} < 1, n > N_d; \quad N_d = floor\left[ \frac{2cL}{\pi D_\rho} - \frac{1}{2} \right],$$

$$A_{nm} = \frac{(-1)^{n+m-1}(4n+2)(\rho_i - \rho_e) D_\rho}{(2m-1)(K_{nm}^+ - K_{nm}^-)\pi^2 L^2}, \quad (16)$$

$$K_{nm}^\pm = -\frac{D_\rho}{2} \left( \frac{(2n+1)^2 \pi^2}{4L^2} + \alpha_m^2 \right) \pm \sqrt{\frac{D_\rho^2}{4} \left( \frac{(2n+1)^2 \pi^2}{4L^2} + \alpha_m^2 \right)^2 - c^2 \left( \frac{(2n+1)^2 \pi^2}{4L^2} + \alpha_m^2 \right)},$$

$$\alpha_m = \frac{(2m-1)\pi}{2H}.$$

The floor function gives the largest integer smaller than its argument. The single sum series in (16) is the part of $\rho'$ that is independent of the coordinate $y$, where $N_d$ is an integer such that $\omega_n$ becomes imaginary when $n > N_d$. For meter-length channels and typical fluid properties $c = 300 m/s, D_\rho = 10^{-5} m^2/s$, $N_d = O(10^7)$. The part of the series with $n > N_d$ is purely diffusive (non-oscillatory). The decay rate for this diffusive portion of the series ($n > N_d$) is about $\exp[-c^2 t / D_\rho]$ or $\exp[-2\gamma_n t]$, whichever is smaller. For $c = 300 m/s, D_\rho = 10^{-5} m^2/s$, $\exp[-c^2 t / D_\rho] \approx \exp[-10^9 t]$, which is an extremely fast decay; whilst $\exp[-2\gamma_n t]$ always decays faster than the terms in the part of series with $n \leq N_d$ which decays as $\exp[-\gamma_n t]$. Thus, the purely diffusive part of the series ($n > N_d$) can be neglected in the calculations for large times due to their faster decay.

The double-sum-series in (16) contains the $y-$dependence of the density profile which vanishes on the channel wall. The slowest decaying mode in the double-sum-series is the $(n, m) = (0, 1)$ mode. A Taylor series expansion for $H/L$ gives, to the leading order,

$$K_{01}^+ = -\frac{c^2}{D_\rho^2}, \tag{17}$$

$$K_{01}^- = -\frac{5.81 D_\rho}{H^2}. \tag{18}$$

$K_{01}^+$ is the aeroacoustic mode, which decays with a rate of $\exp[-c^2 t / D_\rho]$, independent of the length and the gap of the channel; and as shown above, $\exp[-c^2 t / D_\rho]$ decays very fast as $\exp[-10^9 t]$. $K_{01}^-$ is the highly damped viscous mode for narrow channels. For example, for a micron-size gap, $H = 10^{-6} m$ and $D_\rho = 10^{-5} m^2/s$, $K_{01}^- = -5.81 \times 10^7 /s$. Thus, for a micron-size gap, after a very short-time of the order $O(10^{-7} s)$, the double-sum-series in (16) quickly decays to zero and the density relaxes to a profile uniform over the channel cross-section with the density perturbation given by the single-sum series. In other words, the acoustic waves quickly approach the plane wave solution in a narrow channel. This fast relaxation of the density in the transverse direction is due to strong transverse viscous diffusion (strong transverse damping effect) in narrow channels. The transversely relaxed density perturbation is given by

$$\rho'(r,x,t) = \frac{4(\rho_i - \rho_e)}{\pi} \sum_{n=0}^{N_d} \frac{(-1)^n}{2n+1} \exp[-\gamma_n t] \cos \omega_n t \cos \frac{(2n+1)\pi x}{2L} \quad (19)$$

after neglecting the fast decaying purely diffusive modes. The density solution (19) is independent of the coordinate $y$ as well as the gap $H$.

Equation (19) shows that in narrow channels, any density variation in the cross-sectional direction is smoothed out very quickly, while viscous damping of the longitudinal variation of density is extremely slow, on the diffusion timescale $D_\rho/L^2$, since $\gamma_n \approx D_\rho/L^2$. The density perturbation solution (19) represents a damped standing acoustic wave in the channel, with a density node at the exit and antinode at the symmetry line $x=0$.

## 4. The velocity field

For the plane wave solution of the density, eqn. (19), the irrotational velocity obtained from the continuity equation (11) becomes $\mathbf{v}'_{IR} = v'_{x,IR}(x,t)\mathbf{e}_x$ which is given by

$$\begin{aligned}
v'_{x,IR}(x,t) &= RE \sum_{n=0}^{N_d} \hat{v}^{(n)}_{x,IR} \exp[\Omega_n t] \sin \frac{(2n+1)\pi x}{2L} \\
&= \frac{8L}{\pi^2} \frac{\rho_i - \rho_e}{\rho_e} \sum_{n=0}^{N_d} \frac{(-1)^n}{(2n+1)^2} \exp[-\gamma_n t][\gamma_n \cos \omega_n t + \omega_n \sin \omega_n t] \sin \frac{(2n+1)\pi x}{2L}
\end{aligned} \quad (20)$$

where

$$\hat{v}^{(n)}_{x,IR} = -\frac{8L}{\pi^2} \frac{\rho_i - \rho_e}{\rho_e} \frac{(-1)^n}{(2n+1)^2} \Omega_n, \quad (21)$$

$$\Omega_n = -\gamma_n + i\omega_n. \quad (22)$$

This irrotational part of the instantaneous velocity is independent of y and $H$. Since $\gamma_n \ll \omega_n$ when $n < N_d$, the instantaneous irrotational velocity $\mathbf{v}'_{IR}$ scales with the speed of sound as $c\,(\rho_i - \rho_e)/\rho_e$, not the viscous velocity scale as in the Poiseuille's law.

The solenoidal velocity $\mathbf{v}'_{RT}$ is two-dimensional,

$$\mathbf{v}'_{RT} = \mathbf{e}_x v'_{x,RT}(x,y,t) + \mathbf{e}_y v'_{y,RT}(x,y,t), \tag{23}$$

where $\mathbf{e}_x, \mathbf{e}_y$ are the unit base vectors in the $x, y$ directions, respectively. A stream-function $\psi(x,y,t)$ can be introduced such that

$$v'_{x,RT} = \frac{\partial \psi}{\partial y},\; v'_{y,RT} = -\frac{\partial \psi}{\partial x}. \tag{24}$$

Thus, the vorticity equation becomes

$$\frac{\partial}{\partial t}\nabla^2 \psi = \nu_e \nabla^4 \psi. \tag{25}$$

The no-penetration and the no-slip boundary condition on the wall for the overall velocity $\mathbf{v}'$ becomes

$$y = \pm H : \frac{\partial \psi}{\partial y} = -v'_{x,IR};\; \frac{\partial \psi}{\partial x} = 0. \tag{26}$$

Since the flow is symmetric about the centerline $y = 0$, $v'_{x,RT}$ must be an even function of $y$. This requires $\psi$ to be an odd function of $y$. The solenoidal velocity is driven by the no-slip condition imposed on the overall velocity $\mathbf{v}'_{RT}$, with the stream-function given by

$$\psi(x,y,t) = -RE\sum_{n=0}^{N_d} \frac{\hat{V}^{(n)}_{x,IR}}{F_n(H)}\exp[\Omega_n t]\left[\sinh(\sigma_n^+ y) - \frac{\sinh(\sigma_n^+ H)}{\sinh(\eta_n^+ H)}\sinh(\eta_n^+ y)\right]\sin\frac{(2n+1)\pi x}{2L}, \tag{27}$$

where,

$$\sigma_n^+ = \frac{(2n+1)\pi}{2L}, \eta_n^+ = \sqrt{\frac{(2n+1)^2 \pi^2}{4L^2} + \frac{\Omega_n}{v_e}},$$

$$F_n(y) = \sigma_n^+ \cosh(\sigma_n^+ y) - \eta_n^+ \frac{\sinh(\sigma_n^+ H)}{\sinh(\eta_n^+ H)} \cosh(\eta_n^+ y). \tag{28}$$

The solenoidal velocity components are

$$v'_{x,RT}(x,y,t) = RE \sum_{n=0}^{N_d} V_n(x,t) \left[ N \cosh(Ny) - \frac{\sinh(NH)}{\sinh(MH)} M \cosh(My) \right],$$

$$v'_{y,RT}(x,y,t) = -RE \sum_{n=0}^{N_d} N W_n(x,t) \left[ \sinh(Ny) - \frac{\sinh(NH)}{\sinh(MH)} \sinh(My) \right], \tag{29}$$

with

$$V_n(x,t) = -\frac{\hat{v}_{x,IR}^{(n)}}{F_n(H)} \exp[\Omega_n t] \sin \frac{(2n+1)\pi x}{2L},$$

$$W_n(x,t) = -\frac{\hat{v}_{x,IR}^{(n)}}{F_n(H)} \exp[\Omega_n t] \cos \frac{(2n+1)\pi x}{2L},$$

$$N(n) = \sigma_n^+ = \frac{(2n+1)\pi}{2L}, \tag{30}$$

$$M(n) = \eta_n^+ = \sqrt{N^2 + \frac{\Omega_n}{v_e}}.$$

In the small gap limit, the stream-wise solenoidal velocity component is given by

$$v'_{x,RT} = \frac{v'_{x,IR}}{2}\left(1 - 3\frac{y^2}{H^2}\right) + O(H^2) = \frac{1}{2}\left(1 - 3\frac{y^2}{H^2}\right) RE \sum_{n=0}^{N_d} \hat{v}_{x,IR}^{(n)} \exp[\Omega_n t] \sin \frac{(2n+1)\pi x}{2L} + O\left((H/L)^2\right). \tag{31}$$

Thus, in the small gap limit, the stream-wise solenoidal velocity becomes parabolic; however, it slips on the wall so that the overall velocity satisfies the no-slip condition. Clearly, the volumetric flow rate due to the solenoidal velocity on any cross-section is zero,

$$Q_{RT} = \int_{-H}^{H} v'_{x,RT} dy = \frac{v'_{x,IR}}{2} \int_{-H}^{H} \left(1 - 3\frac{y^2}{H^2}\right) dy = 0. \tag{32}$$

This result is general, regardless of the channel gap size, as the volume integration of the incompressibility condition $\nabla \cdot \mathbf{v}'_{RT} = 0$ between $x = 0$ and $x$ gives

$$0 = \int \nabla \cdot \mathbf{v}'_{RT}\, dv = \oiint_{boundary} \mathbf{n} \cdot \mathbf{v}'_{RT}\, da = Q_{RT}(x) - Q_{RT}(0) = Q_{RT}(x), \tag{33}$$

after applying the no-penetration condition on the channel wall and the symmetry condition at $x = 0$. Thus, the solenoidal velocity $\mathbf{v}'_{RT}$ makes no contribution to the mass flow rate and its sole role is to enforce the no-slip condition on the channel wall for the overall velocity.

In the small gap limit, the overall stream-wise velocity is also parabolic,

$$\begin{aligned} \mathrm{v}'_x &= \mathrm{v}'_{x,IR} + \mathrm{v}'_{x,RT} = \frac{3}{2}\left(1 - \frac{y^2}{H^2}\right)\mathrm{v}'_{x,IR} \\ &= \frac{12L}{\pi^2}\frac{\rho_i - \rho_e}{\rho_e}\left(1 - \frac{y^2}{H^2}\right)\sum_{n=0}^{N_d} \frac{(-1)^n}{(2n+1)^2} \exp[-\gamma_n t]\left[\gamma_n \cos(\omega_n t) + \omega_n \sin(\omega_n t)\right] \sin\frac{(2n+1)\pi x}{2L} \end{aligned} \tag{34}$$

While the overall stream-wise velocity is parabolic in the small gap limit, it differs from the plane Poiseuille flow solution

$$\mathrm{v}'_x = -\frac{H^2}{2\mu}\frac{dp}{dx}\left(1 - \frac{y^2}{H^2}\right) \tag{35}$$

in at least two significant aspects:

(i)    The centerline velocity is independent of the gap; whilst the Poiseuille flow solution (35) has a centerline velocity proportional to $H^2$;

(ii)    The velocity does not scale with the inverse of viscosity as in the Poiseuille flow solution (35).

The work above reveals that there are two flow mechanisms for a drainage flow: the compressible part of the flow is driven by the fluid's volumetric expansion, and it is irrotational and it slips on the channel wall; whilst the incompressible part of the flow is driven by the a slip velocity on the channel wall that makes the overall velocity satisfying the no-slip condition. However, the incompressible part of the flow generates no net mass flow rate as shown by eqn. (32).

Fig. 2 shows a sequence of the instantaneous streamline plots over one period for the solenoidal velocity $\mathbf{v}'_{RT}$ for a channel with a gap of $20\mu m$ ( $H = 10\mu m$ ), and half-length of $L = 1m$. Fluid property values are listed in the Appendix. In the plots, only the top-half of the channel is shown and the coordinates $x, y$ are made dimensionless by $L, H$, respectively, so that they run from 0 to 1 and 0 to 1/2, respectively in the plots. However, the velocity values are dimensional, as indicated by the color bar. The period of oscillation is $T = 0.006849s$ and the streamlines plots starts at the time instant of $t_0 = 10^4 T = 68.49s$. Eight time instants are shown, with an increment of $T/8$ starting from $t_0$. The solenoidal velocity slips on the channel wall and for the first half of the period, the slip velocity is in the negative x-direction; whilst for the second half, it is in the positive x-direction. The obvious patterns of these instantaneous streamlines is that the solenoidal flow near the wall and that around the central region are always in the opposite directions, resulting in a zero mass flow rate over any cross-section as shown analytically by eqn. (41).

Fig. 3 shows the corresponding instantaneous streamlines for the overall velocity $\mathbf{v}'$ over the same period. A standout feature of these streamlines is that the overall velocity satisfies the no-slip condition; and except for the short times around the start and the end of the period, significant flows occur only near the central region of the channel. For the first half period, fluid moves out of the channel; whilst for the second half, fluid is sucked back into the channel.

Because of the decay of the velocity in time, there is a net amount of fluid produced during one period as more fluid moves out than the amount sucked back in. This net production will be determined analytically in the next section. Also noticeable is that the instantaneous velocity is relatively large, sine it scales with the speed of sound as alluded to earlier. The period-averaged velocity, however, is much smaller, as will be shown in the next section. Finally, we emphasize that when the property values are varied, these flow patterns remain qualitatively the same for narrow channels.

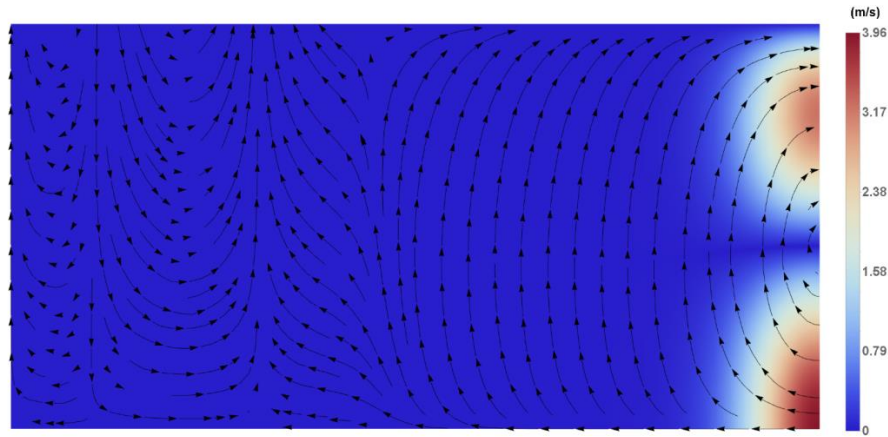

(a) $t_0$

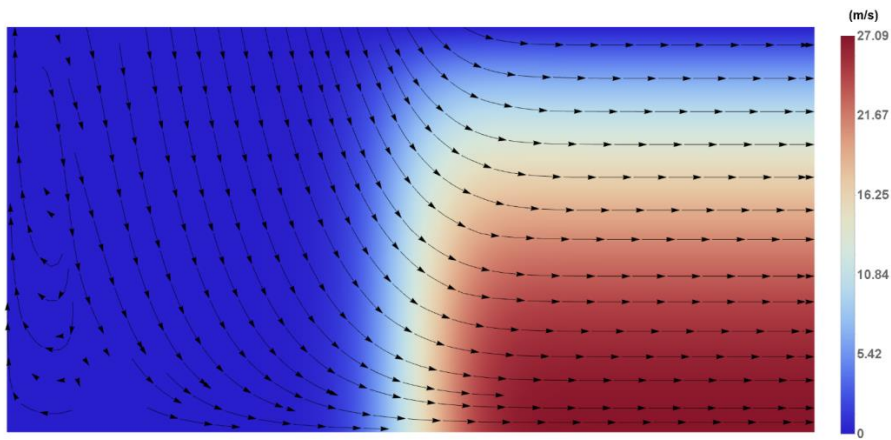

(b) $t_0 + T/8$

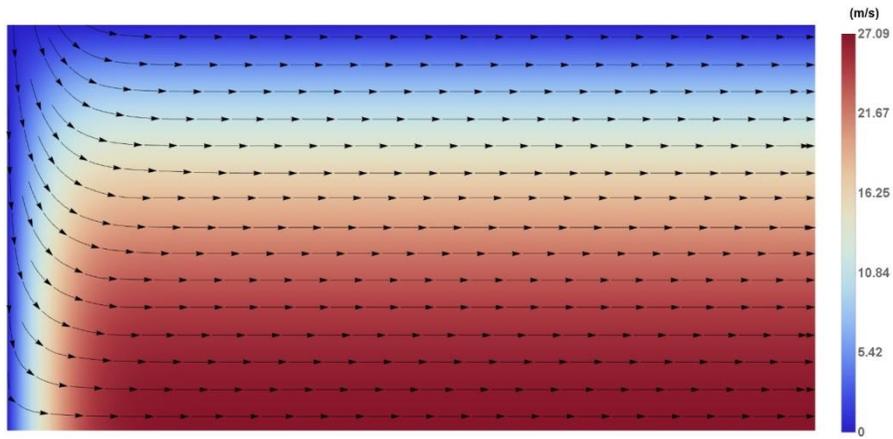

(c) $t_0 + T/4$

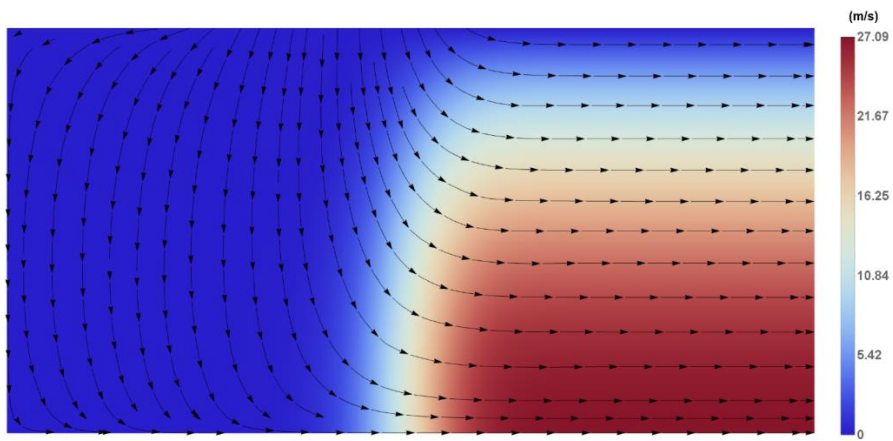

(d) $t_0 + 3T/8$

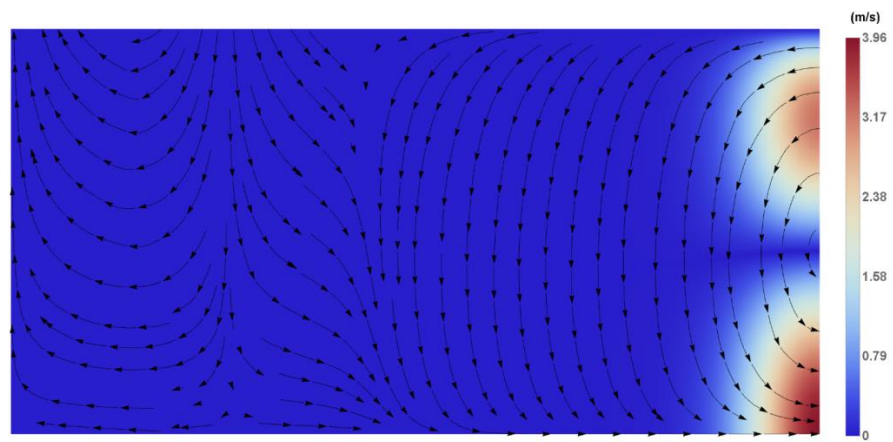

(e) $t_0 + T/4$

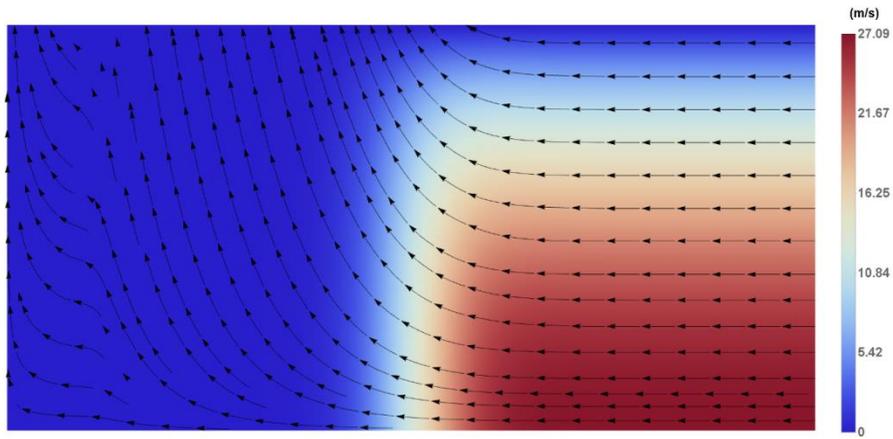

(f) $t_0 + 5T/8$

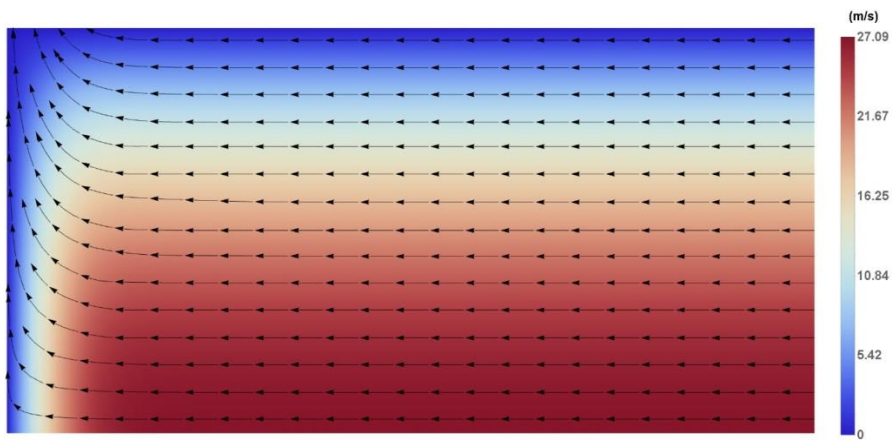

(g) $t_0 + 3T/4$

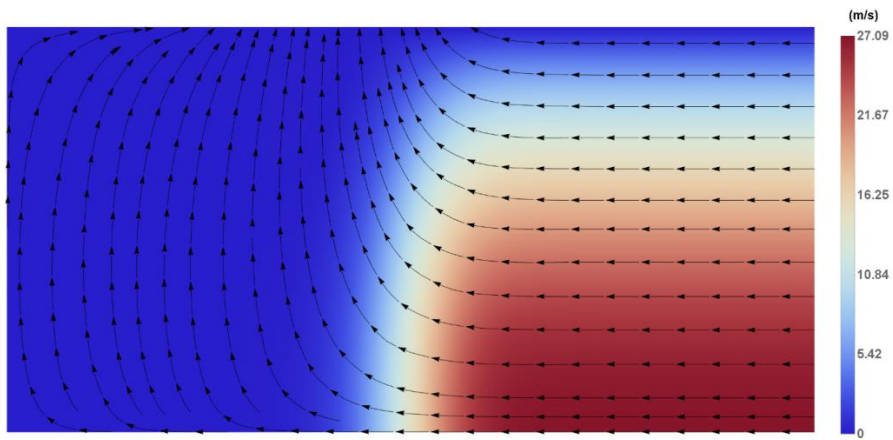

(h) $t_0 + 7T/8$

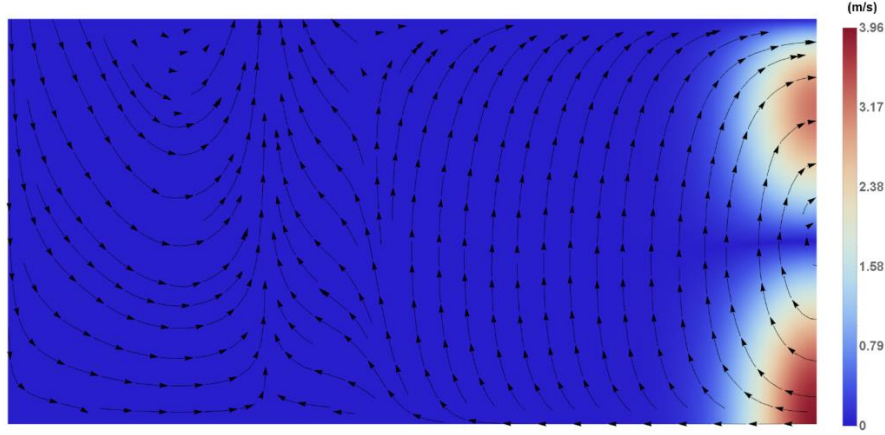

(i)  $t_0 + T$

**Fig. 3** Instantaneous streamlines for the overall velocity over one period.  Only the upper-half of the channel is shown. The velocity satisfies no-slip on the wall. Fluid production is achieved by the decay of the amplitude of oscillation.

## 5. The large-timescale diffusive behavior and the drainage rate

The instantaneous velocity is a damped wave.  The slowest decaying component has a period of the oscillation of $T = 4L/c$, which is about $O(10^{-2}\,s)$ for meter-long channels and it is even shorter for centimeter-long channels.  This is a quite short time compared to the timescale of interest in most of applications.  For sampling times much greater than the period, the behavior of the system is diffusive, even though its short-timescale behavior is ballistic with the instantaneous velocity scaling with $c\,(\rho_i - \rho_e)/\rho_e$.  The large-timescale behavior of the system can be obtained by averaging the instantaneous field over the period.  For example, for the irrotational velocity $v'_{x,IR}(x,t)$, its period-averaged value, denoted by an overbar, is given by

$$\overline{v'_{x,IR}(x,t)} = \frac{1}{T}\int_{t}^{t+T} v'_{x,IR}(x,\xi)d\xi = \frac{8L}{\pi^2}\frac{\rho_i - \rho_e}{\rho_e}\sum_{n=0}^{N_d}\frac{(-1)^n}{(2n+1)^2}\frac{\exp[-\gamma_n t]-\exp[-\gamma_n(t+T)]}{T}\sin\frac{(2n+1)\pi x}{2L},$$

(36)

where $t = kT, k \gg 1$ is an integer. For small $T \ll t$,

$$\frac{\exp[-\gamma_n t] - \exp[-\gamma_n (t+T)]}{T} \approx -\frac{d}{dt}\exp[-\gamma_n t] = \gamma_n \exp[-\gamma_n t].$$

Thus,

$$\overline{v'_{x,IR}(x,t)} = \frac{\rho_i - \rho_e}{\rho_e} \frac{D_\rho}{L} \sum_{n=0}^{N_d} (-1)^n \exp[-\gamma_n t] \sin\frac{(2n+1)\pi x}{2L}. \quad (37)$$

Eqn. (37) clearly shows that the period-averaged irrotational velocity is diffusive, and it scales with the diffusive characteristic velocity $D_\rho / L$. This is very similar to the motion of a Brownian particle in a liquid which shows ballistic behavior in short-timescale and diffusive behavior in large-timescale (Huang et al. 2011).

The drainage rate is the mass flow rate calculated from the period-averaged irrotational velocity at the exit, as the solenoidal velocity produces no mass flow rate at any cross-section (eqn. (32)). Thus, if the channel width is $W$, from equation (37) we have the drainage rate

$$\overline{\dot{m}_e(t)} = \rho_e \overline{v'_{x,IR}(L,t)}(2HW) = 2(\rho_i - \rho_e)\frac{D_\rho}{L} HW \sum_{n=0}^{N_d} \exp\left[-\frac{D_\rho (2n+1)^2 \pi^2 t}{8L^2}\right]. \quad (38)$$

The channel cross-sectional area averaged exit velocity is simply

$$V_e(t) = \overline{v'_{x,IR}(L,t)} = \frac{\rho_i - \rho_e}{\rho_e} \frac{D_\rho}{L} \sum_{n=0}^{N_d} \exp\left[-\frac{D_\rho (2n+1)^2 \pi^2 t}{8L^2}\right]. \quad (39)$$

The same result can be obtained by applying equation (6) and the density solution (19) to the instantaneous mass flow rate at the exit

$$\dot{m}_e(t) = -\int_{Tube} \frac{\partial \rho'}{\partial t} dv = \frac{16 HWL(\rho_i - \rho_e)}{\pi^2} \sum_{n=0}^{N_d} \frac{\exp[-\gamma_n t]}{(2n+1)^2} (\gamma_n \cos\omega_n t + \omega_n \sin\omega_n t). \quad (40)$$

Equation (51) shows that the instantaneous mass flow rate experiences oscillations, with production greater than intake in any given period due to the damping of the oscillation. Thus, when viewed in the short-timescale, fluid is produced in a huff-n-puff process. When the instantaneous mass flow rate (40) is averaged over a period, it produces the same result for the drainage rate $\overline{\dot{m}_e(t)}$ as given by eqn. (38).

A standout feature of the drainage rate formula (38) is that it scales linearly with the channel gap $H$ instead of $H^3$ as given by the Poiseuille's law. Thus, (38) is a slip-like mass flow rate, despite the corresponding velocity profile satisfying the no-slip condition on the channel wall. The drainage rate can be very high in short-times, decreasing diffusively as $t^{-1/2}$ as time is increased. The drainage speed (39) is independent of the channel gap vs. the $H^2$ dependence from the Poiseuille's law. Furthermore, the drainage rate and the drainage speed are proportional to the fluid kinematic viscosity via the diffusion coefficient $D_\rho$, in stark contrast to the inverse proportionality to the fluid viscosity provided by the Poiseuille's law. Altogether, it is evident that the fluid volumetric expansion driven drainage flow considered here is controlled by an entirely different mechanism than Poiseuille-type of flows.

For any given density drop $\Delta \rho = \rho_i - \rho_e$, the mass of the producible fluid from the channel is fixed; for half of the channel considered in the symmetric drainage problem, this producible fluid mass is $M_L = 2HWL(\rho_i - \rho_e)$. The drainage rate in (38) can be expressed as

$$\overline{\dot{m}_e(t)} = M_L \frac{D_\rho}{L^2} \sum_{n=0}^{N_d} \exp\left[-\frac{D_\rho (2n+1)^2 \pi^2 t}{8L^2}\right]. \tag{41}$$

Here, $M_L D_\rho / L^2 \approx$ (fluid mass to be drained)/(characteristic drainage time), and it is the characteristic production rate for the drainage flow. Therefore, the dimensionless mass flow rate is

$$\dot{m}_D(t_D) = \frac{\overline{\dot{m}_e(t)} L^2}{M_L D_\rho} = \sum_{n=0}^{N_d} \exp\left[-\frac{(2n+1)^2 \pi^2}{8} t_D\right], \tag{42}$$

where the dimensionless time $t_D = D_\rho t / L^2$. When $t_D = 7.5$, $\dot{m}_D = 10^{-4}$. Thus, the drain-out time is

$$t_{D,drain-out} = 7.5, \tag{43}$$

or

$$t_{d,drian-out} = 7.5 \frac{L^2}{D_\rho}. \tag{44}$$

The constant becomes 3.75 if drain-out is defined as $\dot{m}_D = 10^{-2}$. The Poiseuille flow theory, on the other hand, gives a drain-out time proportional to the kinematic viscosity, $t_d \propto \nu_e$, instead of the inverse of the kinematic viscosity as shown by (44). The single curve of $\dot{m}_D(t_D)$ is plotted in Fig. 4 on a log-log scale. The mass flow rate decays diffusively with the power $t_D^{-1/2}$ before exponential decay sets in.

The drainage rate (38) also applies to the case of a half-closed channel in which the symmetry line at $x = 0$ is replaced by a sealed end. This is due to the fact that the density solution for this situation is exactly the same as the symmetric drainage problem, although the velocity fields are different.

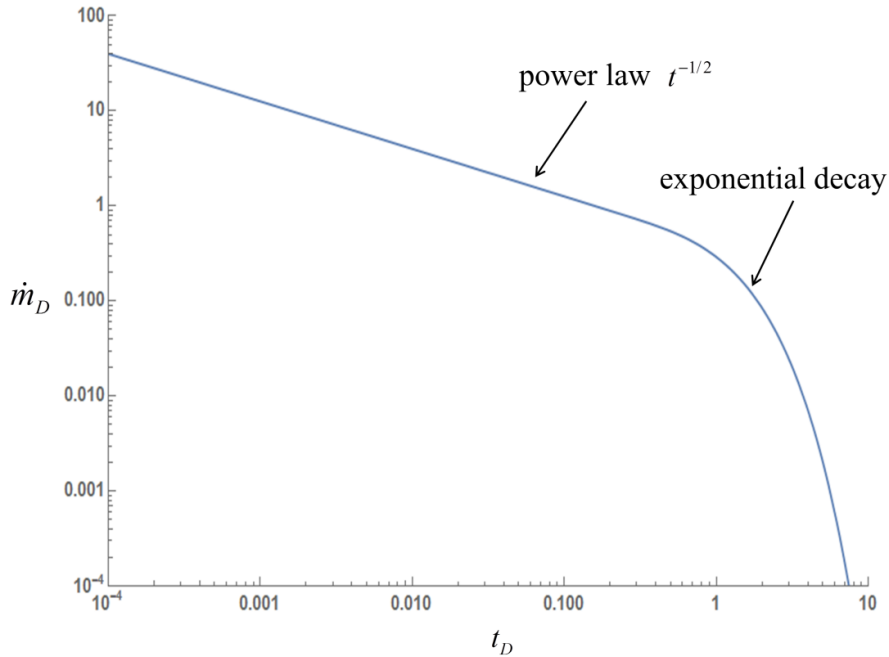

**Fig. 4** Log-log plot of the dimensionless drainage rate vs dimensionless time.

## 6. Conclusions

Symmetric drainage flow of a compressed fluid from a narrow channel is studied on the basis of linearized compressible Navier-Stokes equations with no-slip condition. The solution is valid for small Mach number flows which allows linearization of the compressible Navier-Stokes equations. The solution holds even in the limit of very small channel gap, so long as the Navier-Stokes equations can be used. It is found that the no-slip flow studied here exhibits a time-dependent slip-like mass flow rate (drainage rate) which scales linearly with the channel gap instead of the cubic power of the gap predicted from Poiseuille flow solution. The drainage rate formula also applies to drainage flow from a half-sealed microchannel. Furthermore, on the large-timescale, the flow is controlled by the diffusion of the density, causing the drainage rate to be proportional to the fluid's kinematic viscosity, opposite to Poiseuille-type of flow which has a mass flow rate proportional to the inverse of kinematic viscosity. These results are similar to those for drainage flow in circular capillaries.

**Appendix**

Fluid properties used in Fig. 2, 3:

Methane at $80^0 C$, $25 MPa$: from http://www.peacesoftware.de/einigewerte/methan_e.html

$\mu = 1.99 \times 10^{-5} Pa \cdot s$ , $\mu_b = 320\mu = 6.368 \times 10^{-3} Pa \cdot s$ , $\rho_i = 136.78 kg/m^3$ , $c = 584 m/s$ ,

$D_\rho = 4.82 \times 10^{-5} m^2/s$;

We choose: $\rho_e = 132.68 kg/m^3$; $L = 1m$; the period $T = \dfrac{4L}{c} = 0.006849 s$.